\documentclass[%
 aip,
 rsi,
 amsmath,amssymb,
 reprint,%
]{revtex4-1}

\usepackage[dvipdfmx]{graphicx}% Include figure files
\usepackage{dcolumn}% Align table columns on decimal point
\usepackage{bm}% bold math
\usepackage[utf8]{inputenc}
\usepackage[T1]{fontenc}
\usepackage{mathptmx}
\usepackage{color}

\begin{document}

%\preprint{AIP/123-QED}

\title{A novel experimental setup for an oblique impact onto an inclined granular layer}
% Force line breaks with \\

\author{Shinta Takizawa}
 \affiliation{Department of Earth and Environmental Sciences, Nagoya University, Nagoya 464-8601, Japan}%Lines break automatically or can be forced with \\
\author{Ryusei Yamaguchi}%
\affiliation{Technical Center, Nagoya University, Furocho, Chikusa, Nagoya 464-8601, Japan}%

\author{Hiroaki Katsuragi}
\affiliation{Department of Earth and Environmental Sciences, Nagoya University, Nagoya 464-8601, Japan}

\date{\today}% It is always \today, today,

\begin{abstract}
We develop an original apparatus of the granular impact experiment by which the incident angle of the solid projectile and inclination angle of the target granular layer can be systematically varied. Whereas most of the natural cratering events occur on inclined surfaces with various incident angles, there have not been any experiments on oblique impacts on an inclined target surface. To perform systematic impact experiments, a novel experimental apparatus has to be developed. Therefore, we build an apparatus for impact experiments where both the incident angle and the inclination angle can be independently varied. The projectile-injection unit accelerates a plastic ball (6~mm in diameter) up to $v_i\simeq 100$~m~s$^{-1}$ impact velocity. The barrel of the injection unit is made with a three-dimensional printer. The impact dynamics is captured by high-speed cameras to directly measure the impact velocity and incident angle. The rebound dynamics of the projectile (restitution coefficient and rebound angle) is also measured. The final crater shapes are measured using a line-laser profiler mounted on the electric stages. By scanning the surface using this system, a three-dimensional crater shape (height map) can be constructed. From the measured result, we can define and measure the characteristic quantities of the crater. The analyzed result on the restitution dynamics is presented as an example of systematic experiments using the developed system.
\end{abstract}

\maketitle

\section{\label{sec:Introduction}Introduction}
Impact cratering is one of the most frequently occurring events on the surface of celestial bodies. We can easily find numerous craters on the surfaces of planets, satellites, asteroids, and comets. The size distribution and morphology of craters provide useful clues about the surface processes occurring on the surfaces of these bodies. A large number of studies on impact cratering have been performed to date~\cite{Melosh:1989,Melosh:2011,Osinski:2013}. While the majority of impact craters possess more or less circular shapes, some of them show peculiar shapes. For example, asymmetric craters have been found on Mars~\cite{Herrick:2006}, Earth~\cite{Kenkmann:2009}, the Moon~\cite{Neish:2014}, and the asteroid Vesta~\cite{Krohn:2014}. In this study, we focused on the development of an experimental apparatus that can vary the impact geometry (impact angle and target inclination) to simulate asymmetric cratering on the surface of a granular (regolith) layer.

\begin{figure}
\includegraphics[clip,width=0.8\linewidth]{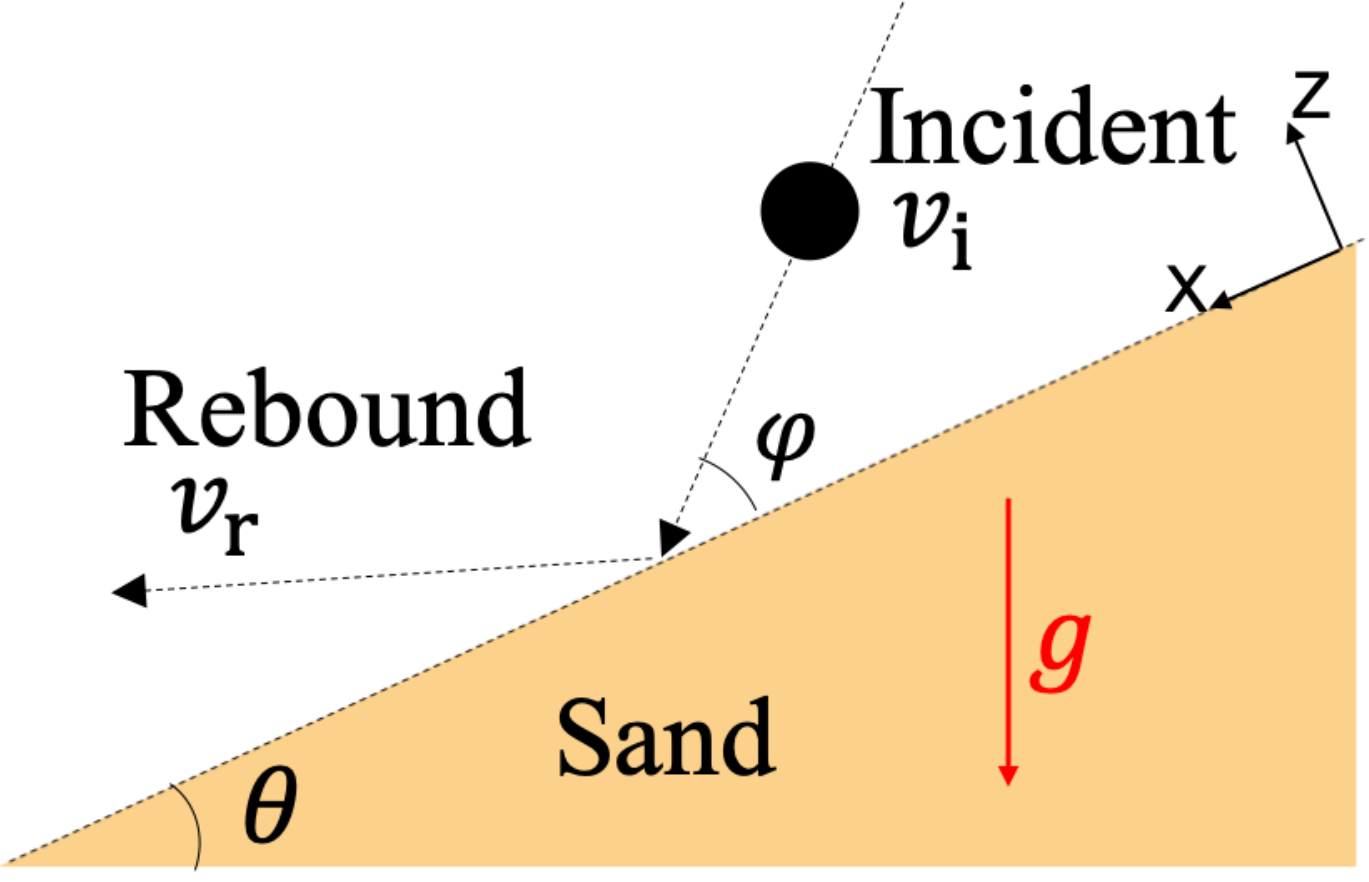}% Here is how to import EPS art
\caption{\label{fig:F0-setup} Definitions of the angles and coordinate system of the experimental setup. The inclination angle $\theta$ and incident angle $\varphi$ are defined by the horizontal and target-surface planes, respectively. The $X$ direction is taken along the direction of slope and the $Z$ direction corresponds to the normal to the target surface. The $XY$ plane corresponds to the target surface. Modified with permission from Icarus 335, 113409 (2020). Copyright 2020 with Elsevier.}
\end{figure}

In general, the effects of oblique impact and target inclination can significantly modify an isotropic transient crater shape, resulting in asymmetric cratering. Therefore, we considered two principal parameters: incident angle $\varphi$ and inclination angle of the target $\theta$ (Fig.~\ref{fig:F0-setup}). However, no experimental studies have been carried out on oblique impacts on an inclined target (sloped terrain). Each effect has been studied separately in the prior studies. 

With regard to the oblique impact effect, Gault and Wedkind (1978) performed a systematic experiment with a solid projectile impacting a horizontal granular layer~\cite{Gault:1978}. According to their results, a rebound/ricochet of the impactor can be observed in shallow impacts ($\varphi<30^{\circ}$). In addition, they found that only a very shallow impact ($\varphi<15^{\circ}$) can result in an asymmetric crater formation, which is why the majority of impact craters have a circular (symmetric) shape. Based on these experimental results, the effects of oblique impacts on crater shape has not been regarded as an important issue. However, only the horizontal target layer was used in this study. 

The effects of target inclination were investigated in two relatively recent studies~\cite{Aschauer:2017,Hayashi:2017}, where the inclination angle $\theta$ of the target granular layer was varied and the projectile was vertically impinged onto the inclined granular surface. Although the ranges of impact velocity were different in the two experiments~(Refs.~\onlinecite{Aschauer:2017,Hayashi:2017}), they observed similar results. As $\theta$ increased, the scale of the landslide triggered by the impact became large and tended to erase the transient crater cavity. These studies independently confirmed the importance of $\theta$ for properly evaluating the crater formation process, including modification by asymmetric collapse. However, both of these experiments used only a vertical impact. In particular, the incident angle $\varphi$ was not systematically varied in these experiments. 

Low-speed (impact velocity $v_i \sim 10^0$~m~s$^{-1}$) granular experiments have long been performed using very simple free-fall drop mechanism. For instance, simple scaling laws for low-speed granular impact cratering have been developed based on very simple experiments~\cite{Walsh:2003,Uehara:2003}. Additionally, the penetration dynamics of a projectile into a granular layer has  also been extensively studied~\cite{Katsuragi:2007,Goldman:2008,Seguin:2009,Katsuragi:2013,Katsuragi:2013,Clark:2014}. In these granular-physics studies, the vertical impact of the projectile onto a horizontal granular target has been focused to derive the fundamental features of the granular impact dynamics. Recent reviews on granular impact cratering and penetration dynamics can be found in Refs.~\onlinecite{RuizSuarez:2013,katsuragi:2016,vanderMeer:2017}. The oblique impact onto an inclined target is an important next step also in terms of fundamental granular impact studies. 

The effect of gravity has to be properly considered to mimic various astronomical impact conditions. However, low-gravity impact experiments have been a challenging problem. For instance, laboratory-scale drop tower systems have been developed~\cite{blum:2014,Sunday:2016}. Using these systems, low-gravity granular impact experiments have been performed recently~\cite{Katsuragi:2018,Murdoch:2017}. Atwood machine mechanisms have also been used in some experiments to partly reduce the effects of gravity~\cite{Goldman:2008,Altshuler:2014}. And more directly, space-shuttle flights~\cite{Colwell:1999,Colwell:2003} and parabolic flights~\cite{Colwell:2008} have been used to perform low-gravity granular impact experiments. Recently, the experimental data obtained by these flights have been combined and systematically analyzed~\cite{Brisset:2018}. Although the experimental apparatuses developed for these flights are very sophisticated, there are some limitations with these flight experiments. For instance, the final crater shape made by the impact cannot be measured, and the impact velocity range is limited ($\lesssim$ 1~m~s$^{-1}$). Actually, such a low-speed impact is appropriate for mimicking spacecraft touchdown on the surface of small asteroids~\cite{Brisset:2018}. To examine the impact crater shapes and scaling laws, however, faster impacts should also be studied.  While the effect of target inclination has been experimentally investigated with the drop-tower experiment~\cite{Hofmann:2017}, it is still difficult to vary both the impact angle and the target inclination under low-gravity conditions.

From the technical point of view, the development of a low-gravity impact system is very challenging and interesting. However, it is not easy to develop a setup of an oblique impact onto an inclined target with controllable angles under low-gravity conditions. First, we had to establish an experimental technique for an oblique impact onto an inclined surface on the ground. Thus, we focused on the development of an experimental apparatus by which $\varphi$ and $\theta$ are independently controlled in the laboratory (under the influence of gravity). The developed system consists of a projectile-injection unit and precise measurement systems for impact dynamics and final crater shape.

\begin{figure}
\includegraphics[clip,width=1\linewidth]{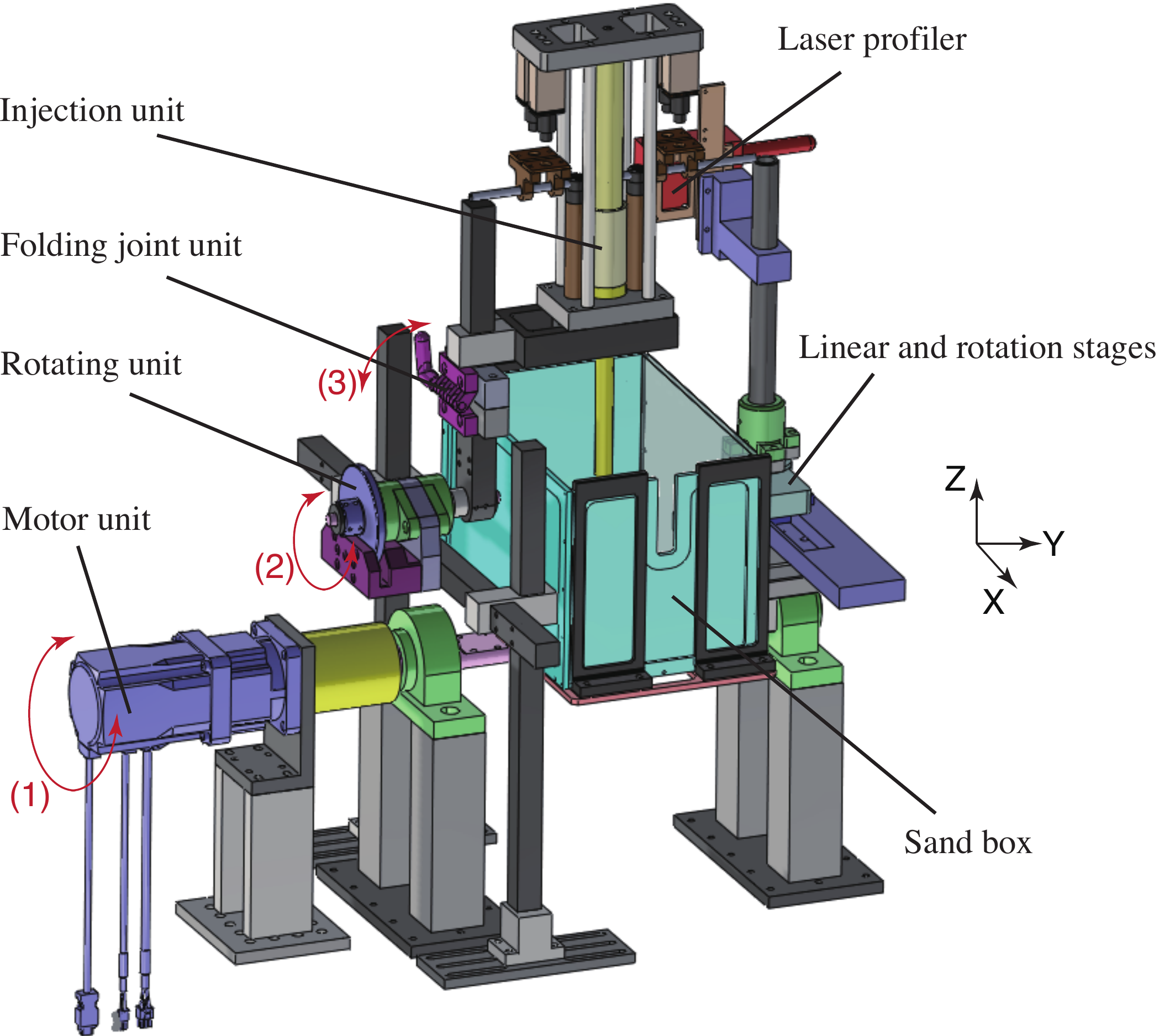}% Here is how to import EPS art
\caption{\label{fig:F1-design} The design of the experimental system. The injection unit can be manually tilted with the rotating unit and laid down at the folding joint unit. The sandbox can be tilted using the stepping motor. The laser profiler is mounted on the linear and rotation stages, which are fixed on the tiltable sandbox system. The laser profiler scans the surface of the target layer.  The curved arrows (1), (2), and (3) indicate, respectively, the rotational directions of the sandbox (and profiler), the injection unit, and the folding joint unit. The common rotational axis of the inclination of target (1) and the injection (2) is the $Y$-axis. The folding unit can fold the injection unit around the $X$-axis.}
\end{figure}

\section{Experimental apparatus}
We developed a system of an injection unit and rotatable sandbox as shown in Fig.~\ref{fig:F1-design}. The injection unit and sandbox container can be tilted manually or with a stepping motor, respectively. A laser profiler is also mounted on the base system (rotatable stage) to measure the surface profile of the target. In the following subsections, details of these units are described.

\subsection{Injection unit}
\begin{figure}
\includegraphics[clip,width=0.8\linewidth]{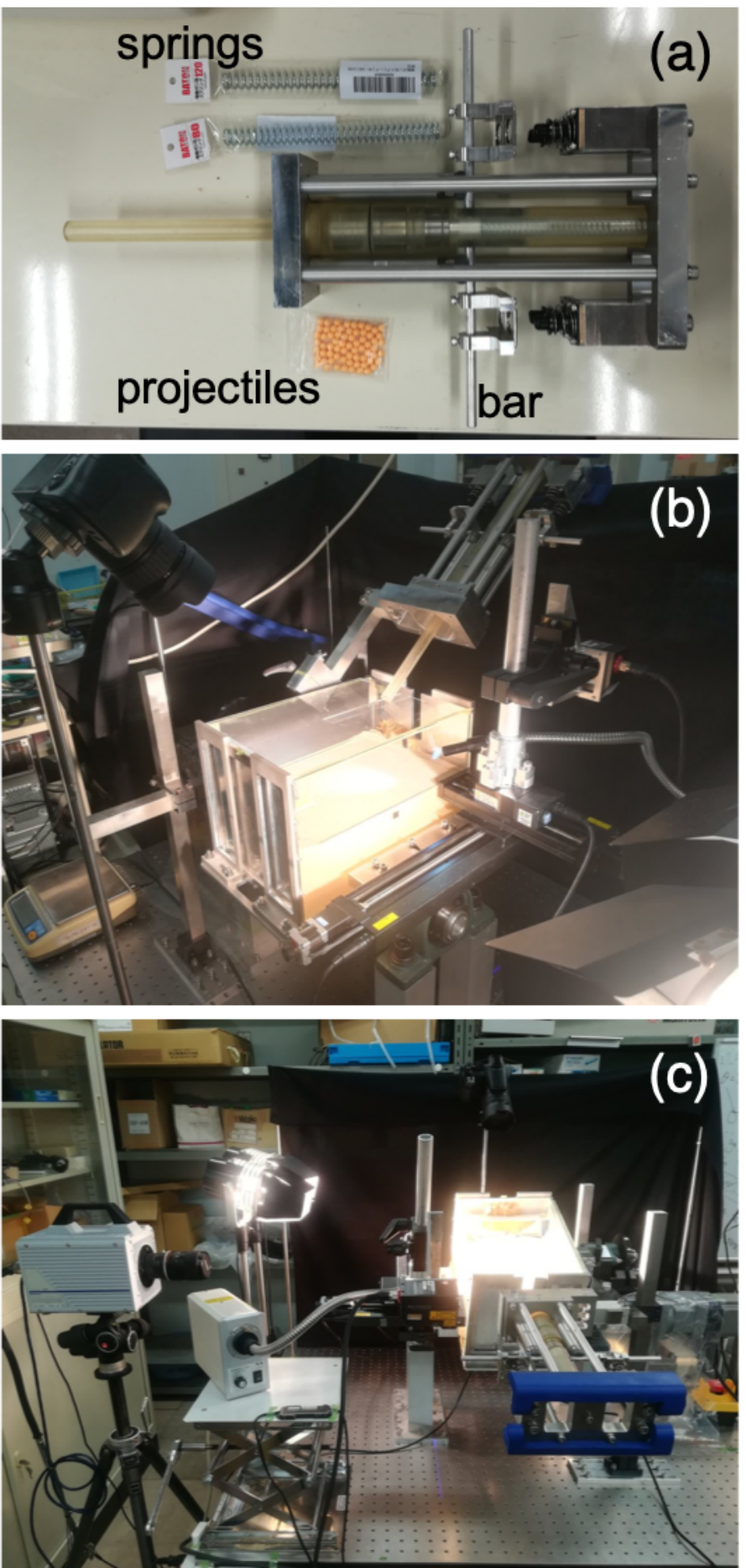}% Here is how to import EPS art
\caption{\label{fig:F2-app} Pictures of the apparatus. (a) The injection unit, shown with springs, units, and projectiles. The barrel of gun was made from an acrylic resin using a 3D printer. The spring is compressed by manually pulling the expanding spring that propels the piston forward, and then the compressed air accelerates the projectile. Using this system, a plastic projectile of 6~mm in diameter can be accelerated up to a speed about 100~m~s$^{-1}$. (b) Typical setup of the oblique impact is shown. The target is illuminated, and the topview camera is placed at the opposite side of the impact. (c) The setup for a very shallow impact angle is shown. The sideview high-speed camera is also visible in this picture. }
\end{figure}

The injection unit is basically made of aluminium and plastic parts; the injection unit with springs and projectiles is shown in Fig.~\ref{fig:F2-app}(a). A spring is embedded in the barrel unit, and by pulling a bar, the spring is compressed by the piston in front of it. The bar is hooked on the push-catch units on both sides. The push-catch units can hold the spring compressed and release it using a toggle mechanism. A plastic ball with a diameter of $D_i=6$~mm and mass of $0.12$~g is placed in the barrel prior to triggering. For testing the projectile material, we chose bullets for toy guns, specifically, a gun used in our recent experiment~\cite{Takizawa:2019}. With this type of projectile, several different density (and same-size) projectiles are commercially available at a low cost. Although the projectile density was fixed in this study, other projectile density data were reported in Ref.~\onlinecite{Takizawa:2019b}. By manually releasing the bar, the spring stretches and compresses the air between the piston and the projectile, and consequently, the compressed air accelerates the projectile. By replacing the spring and/or manually controlling the projectile position in the gun barrel, the projectile speed $v_i$ can be controlled within the range of $10 \leq v_i \leq 100$~m~s$^{-1}$. The position of the projectile in the barrel determines the volume of compressed air, which significantly affects the injection speed. Because $v_i$ is manually controlled, its control accuracy is very limited, and while we can roughly anticipate the projectile speed by manually controlling some factors, the actual value of $v_i$ must be directly measured with a high-speed camera for quantitative analysis.

The barrel is shaped by a three-dimensional (3D) printer (KEYENCE, AGILISTA-3100). The material used to make the model was acrylic resin (AR-M2). The 3D printer enables us to flexibly design an injection unit that is easy to test and correct errors in, but at a low cost. In particular, when we designed the original injection unit, rapid prototyping was necessary for efficient development.  For now, we have made the barrel with only an inner diameter of $6$~mm. However, with the 3D printer, it is easy to make various barrel sizes for projectiles ranging from $5$~mm to $30$~mm in diameter.

The injection unit is held on the base system with a rotation unit that can tilt the injection unit at any angle ($360^{\circ}$ rotatable). Oblique impact setups with (b)~$\varphi\simeq 45^{\circ}$ and (c)~$\varphi \simeq 170^{\circ}$ are shown in Fig.~\ref{fig:F2-app}(b,c). The incident angle is manually controlled, and the angle is measured using an angle meter (SK Niigata seiki Bevel Box BB-180L at a  resolution of 0.1$^{\circ}$). Before and after the impact, the surface profiles were measured to obtain the crater shape by subtracting the before-impact profile from the after-impact profile~(Sec.~\ref{sec:profilometry}). In order to secure the open space for the surface measurement, the injection unit can be laid down using the folding joint unit which can be opened and closed by a lever. (Fig.~\ref{fig:F1-design}). 

The actual impact velocity and incident angle were measured by a high-speed camera (Photron, FASTCAM SA5) with a frame rate of 10,000 frames per second and spatial resolution of $0.18$~mm per pixel. This camera was placed beside the system~(Fig.~\ref{fig:F2-app}(c)), and to precisely measure these impact parameters, the muzzle was placed at least 100~mm away from the target surface. Otherwise, it is impossible to capture a sufficient number of projectile snapshots for $v_i$ measurement with this experimental setup. If a rebound of the projectile was observed, the rebound speed and its angle were also measured by the same high-speed camera. Another high-speed camera (CASIO, Exilim EX-F1) was placed above the target sandbox to acquire a topview of the crater formation dynamics (Fig.~\ref{fig:F2-app}(b)). The frame rate and spatial resolution were set to $300$ frames per second and $0.5$~mm per pixel, respectively. This camera was mainly used to observe the qualitative behavior of the cratering dynamics. These two high-speed cameras were not synchronized. The impact moment (time $t=0$) was identified by each taken movie. For this experimental setup, the frame rate and spatial resolution of the sideview video were determined to precisely measure the impact speed and its angle. For the topview video, 300 frames per second was the maximum frame rate to observe a sufficiently large field of view. However, this frame rate was insufficient to track the projectile's motion. That is why the topview camera was used only for the qualitative analysis. Two halogen lamps (LPL, video light VL-1300/G), and an LED spotlight (Hayashi watch-works, LED light LA-HDF158AS) were used for illumination.

\subsection{Tiltable sandbox target}
We used natural sand (TOYOURA KEISEKI KOGYO, K.K., Toyoura sand) as target material, which has a grain diameter and true density of $D_g=0.1$-$0.3$~mm and $\rho_g=2.6 \times 10^3$~kg~m$^{-3}$, respectively. The angle of repose of Toyoura sand is $34^{\circ}$. A sandbox with aluminium or acrylic walls (inner dimensions: 200~mm$\times$300~mm$\times$200~mm) was mounted on a tiltable table that was rotated by a stepping motor (Oriental motor, AZM98MC-HS100). A transparent acrylic wall was used on one side specifically for collecting high-speed video data. By using a 1/100 reducer, the actual rotational resolution and maximum torque of the motor unit were 0.0036$^{\circ}$ per pulse and 52~Nm, respectively. Sand was poured into the box up to 100~mm thickness. According to Refs.~\onlinecite{Seguin:2008,Nelson:2008}, this system size is sufficiently large to avoid the container-wall effect. The specific size of the container was determined by considering the available system size, reasonable ability of the stepping motor, and so on. Toyoura sand grains were glued to the inner side walls of the container box up to $100$~mm thickness to create frictional wall conditions. Using an acrylic spatula, the surface was flattened and leveled to maintain the target surface parallel to the container bottom wall. Before every impact, the sand layer was stirred and leveled to make a homogeneous target layer. Although the air fluidization of the target granular layer~\cite{Katsuragi:2007} was ideal for making a homogeneous layer, the manual stirring was employed in this study because the sandbox was too large to be fluidized. Surface flatness is crucial to obtain reproducible results on crater morphology. Thus, several surface-flattening methods were tried, and we found that the flattening with a spatula fitted to the sandbox size was the best method of target preparation. The bulk packing fraction of the target was fixed at 0.55. On the side wall, a vertical slit was opened to accommodate the projectile gun for tests with very shallow incident angles (e.g., Fig.~\ref{fig:F2-app}(c)). The rotational axis was defined as the $Y$ direction, and the normal to the sand surface corresponded to the $Z$ direction. Namely, the crater was made on the $XY$ plane (Figs.~\ref{fig:F0-setup} and \ref{fig:F3-laser}).

\subsection{Profilometry system}
\label{sec:profilometry}
\begin{figure}
\includegraphics[clip,width=0.8\linewidth]{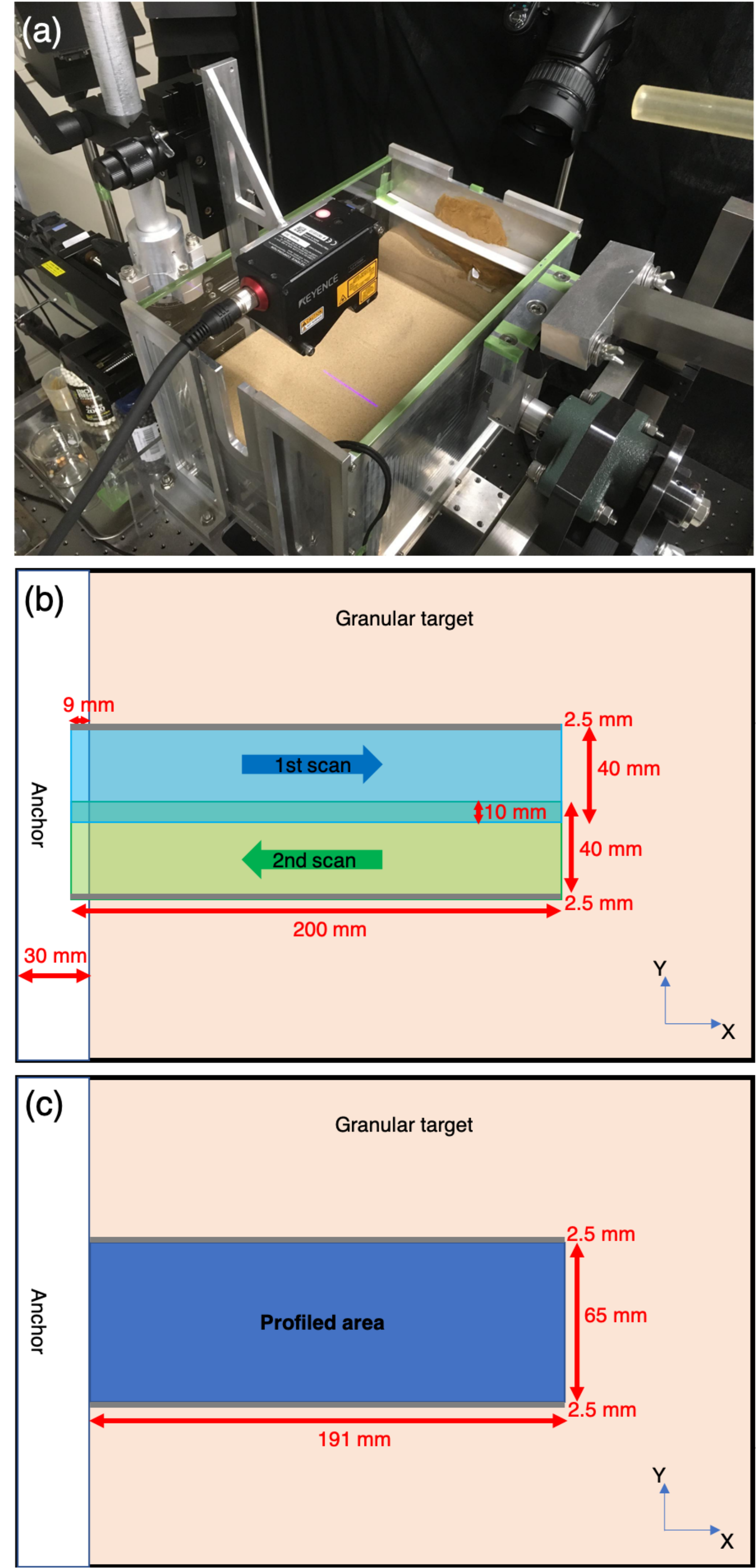}% Here is how to import EPS art
\caption{\label{fig:F3-laser} Laser profilometry system and the region of interest of the surface profile measurement. (a) The laser profiler measuring the surface profile is shown. The purple laser line on the granular surface is visible, and the muzzle of the folded injection unit can also be seen on the right side. Using the linear stages, the laser profiler can move in both the $X$ and $Y$ directions. The rotational stage is used to place the laser profiler at the safety position during the impact. (b) The area scanned by the laser profiler is shown. The round-trip measurement with different $Y$ position enables the wider surface measurement. (c) The final region of interest that can be synthesized from the scanned profiles is presented. The edge $2.5$~mm data were not used because the edge data were sometimes unstable.}
\end{figure}

The surface profile of the target sand layer was measured by a line laser profiler (KEYENCE, LJ-V7080). The laser profiler can take cross-sectional profiles of roughly 40~mm linear segments with a horizontal resolution of 50~$\mu$m. In this experimental setup, the laser line was aligned to the $Y$ direction. The vertical resolution of the measured profile was 0.5~$\mu$m. The measurable range of the vertical distance was $80$~mm $\pm 23$~mm, which was sufficiently large for our setup ($6$~mm projectile impact onto a granular layer at $100$~m~s$^{-1}$). The laser profiler was attached to the electronic stages (COMS, 200~mm stroke in $X$-direction: PM80B-200X, 100~mm stroke in $Y$-direction: PM80B-100X, and $360^{\circ}$ rotation along $Z$-axis: PS40BB-360R). Because these stages were mounted on the tilting table (sandbox system), the profiler scanned on the XY plane even in a steeply tilted case. A picture of the actual measurement system is shown in Fig.~\ref{fig:F3-laser}(a). The rotation stage was used to move the laser profiler out of the way during testing and back in place to take final measurements. Before and after each impact, the surface profile of the sand layer was measured by scanning the surface. The laser profiler was moved in the $X$ direction at a speed of $5$~mm~s$^{-1}$. The height data were taken every $10$~ms so that the spatial resolution in the $X$ direction became $50$~$\mu$m, which was identical to the spatial resolution in the $Y$ direction. To measure a region wider than 40~mm, the laser profiler was shuttled in displaced regions, as shown in Fig.~\ref{fig:F3-laser}(b). By combining the round-trip data, the surface profile in an area of a 191~mm$\times$65~mm region can be synthesized (Fig.~\ref{fig:F3-laser}(c)). The measurement at the edge of the laser line was sometimes unstable due to limitations of the instruments. Thus, we discarded the edge $2.5$~mm data from the combined profile.

We carefully calibrated and aligned the measurement system to accurately measure the surface profile. Specifically, the angle between the horizontal surface and the profiler's motion axis was measured by scanning the horizontal acrylic plate placed on the sandbox. Using this measurement, we confirmed that the measured height difference between two edges in the longer direction ($300$~mm), $O(10^{-2}\mbox{~mm})$, was much less than the grain diameter, $O(10^{-1}\mbox{~mm})$. Thus, we could avoid the inclination motion between the sandbox and the profiler. The difference between the first and second scans in the overlapped region (10~mm width, shown in Fig.~\ref{fig:F3-laser}(b)) was also in the order of $O(10^{-2}\mbox{~mm})$. The simple average was sufficient to join the data of two scans. By measuring the anchor (Fig.~\ref{fig:F3-laser}(b,c)), whose position was known, the length could be calibrated. We confirmed that all the measured results obtained using this protocol were reasonably stable. Thus, we conclude that the system is accurate enough to quantify crater morphology, at least on a grain-sized scale. The height difference before and after the impact on the $XY$ plane (denoted by $\delta Z$) was used for the analysis of crater morphology. A similar laser profilometry system was formerly developed in order to characterize droplet impact cratering~\cite{Katsuragi:2010,Katsuragi:2011}. By replacing a laser profiler with the new one and combining two electric stages, the measurement accuracy, area, and time were significantly improved in the current system. 

\section{Experimental Results}
Using the developed system, we performed a set of systematic impact experiments in which the incident angle $\varphi$ and target inclination angle $\theta$ were independently varied. A detailed analysis of crater morphology using a dimensional analysis method was reported in Ref.~\onlinecite{Takizawa:2019b}. Here, we show some example data to demonstrate the capability of the developed system. Most of the data used in this paper are identical to those used in Ref.~\onlinecite{Takizawa:2019b}; however, the example images and analyzed data shown in this paper are not presented in that study. In addition, the data on projectile rebounds were not analyzed in that study. In this paper, we analyzed the impact events where projectile rebounds were observed. Specifically, we analyzed the data from $52$ impacts and rebounds ($\theta=0^{\circ}$:11,$\theta=10^{\circ}$:8,$\theta=20^{\circ}$:25,$\theta=30^{\circ}$:18).

\begin{figure}
\includegraphics[clip,width=0.9\linewidth]{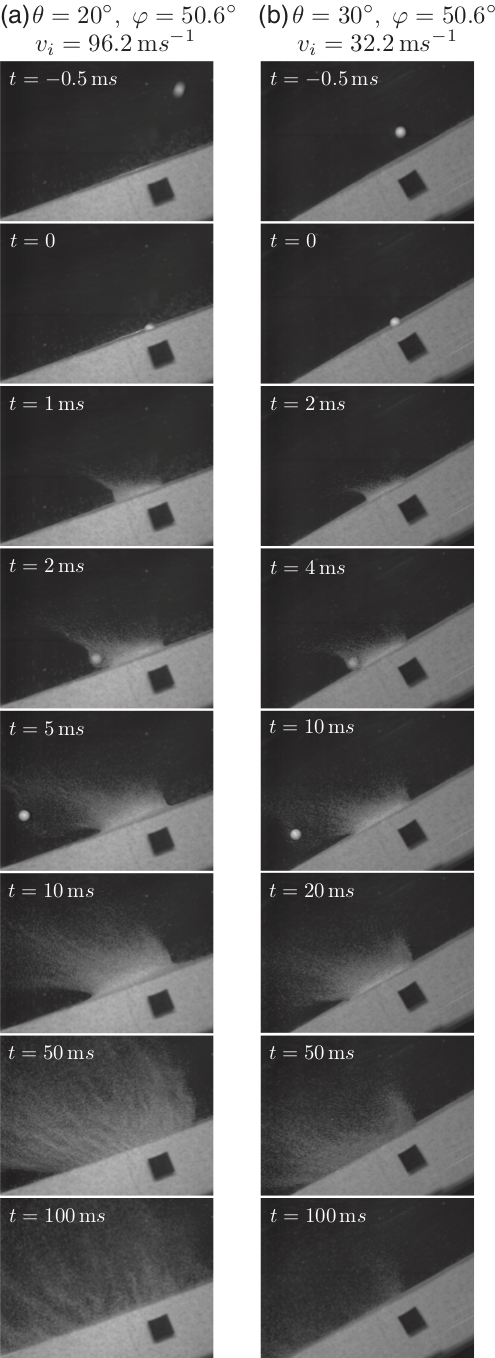}% Here is how to import EPS art
\caption{\label{fig:F4-photron} Sideview images of the impacts. Experimental conditions are described above the images. The main differences between (a) and (b) are inclination angle and impact velocity. Asymmetric ejecta splashing and rebound of the projectile can be confirmed in both cases.}
\end{figure}

\subsection{Impact dynamics}\label{sec:impact_dynamics}
\subsubsection{Sideview and image analysis}
\label{sec:sideview_image}
First, examples of the sideview (high-speed) images  are presented in Fig.~\ref{fig:F4-photron}~((a)~$\theta=20^{\circ}$, $\varphi=50.6^{\circ}$, and $v_i=96.2$~m~s$^{-1}$, and (b)~$\theta=30^{\circ}$, $\varphi=50.6^{\circ}$, and $v_i=32.2$~m~s$^{-1}$). The value of $\theta$ was precisely controlled by using the stepping motor. Therefore, we simply used the $\theta$ value set by the motor controller. Of course, we confirmed that the $\theta$ value was consistent with the image taken by the camera. Using this type of image, we measured the impact velocity $v_i$, impact angle $\varphi$, rebound velocity $v_{\rm re}$, and rebound angle $\varphi_{\rm re}$.

To measure $v_i$, $\varphi$, $v_{\rm re}$, and $\varphi_{\rm re}$, we assumed the projectile motion was restricted on the $XZ$ plane. Specifically, we ignored the motion in the $Y$ direction. Because we aligned the system so that the gun barrel was parallel to the $X$-axis, and the camera's optical axis was perpendicular to the $XZ$ plane, this assumption is reasonable. Although it was difficult to quantify the motion in the $Y$ direction, we always observed the symmetric crater shape (mirror symmetry along the $X$-axis at the crater center). Moreover, any significant motion of the projectile in the $Y$ direction cannot be observed in the topview images, even after the rebound. These are indirect evidences for the negligible $Y$ component in $v_i$, etc.

The position of the projectile was measured by computing the center of mass of the projectile image using ImageJ software. In Fig.~\ref{fig:F4-photron}(a), the projectile image before impact is sightly blurred due to the limited shutter speed, $0.1$~ms (reciprocal of 10,000 frames per second was used to maximize the gain). However, we can analyze the position of the projectile because the center of mass can be reliably computed even for this level of blurred image. Given that this impact speed roughly corresponds to the upper limit achieved by this experimental system, we can safely analyze all the impact data with this frame rate and shutter speed. Indeed, the projectile shape can clearly be confirmed in the case of the slower $v_i$, as shown in Fig.~\ref{fig:F4-photron}(b). 

Using these data, $v_{i}$, $\varphi$, $v_{\rm re}$, and $\varphi_{\rm re}$ were measured using the following procedures. Specifically, the velocities can be computed from the difference of projectile positions in two successive snapshots. The maximum velocities before and after the impact were picked up as $v_i$ and $v_{\rm re}$, respectively. The angles $\varphi$ and $\varphi_{\rm re}$ were measured from the corresponding angles. The errors were estimated from the standard deviations of the data set available to compute these quantities (around five frames). The measured values for the data shown in Fig.~\ref{fig:F4-photron} were (a) $v_i=96.2 \pm 0.4$~m~s$^{-1}$, $\varphi=50.6\pm 0.1^{\circ}$, $v_{\rm re}=0.58 \pm 0.10$~m~s$^{-1}$, $\varphi_{\rm re}= 66\pm 18^{\circ}$ and (b) $v_i=32.2 \pm 0.2$~m~s$^{-1}$, $\varphi=50.6\pm 0.2^{\circ}$, $v_{\rm re}=3 \pm 1$~m~s$^{-1}$, $\varphi_{\rm re}= 80\pm 1^{\circ}$. Because the injection unit was carefully tilted with monitoring of the angle meter, $\varphi$ is reproducibly controlled. The relationships among $\varphi$, $\theta$, $\varphi_{\rm re}$, and the restitution coefficient $e$ are discussed later, in Sec.~\ref{sec:rebound}. 

As can be confirmed in Fig.~\ref{fig:F4-photron}, ejecta splashing driven by the oblique impact shows an asymmetric nature. This asymmetry results in the asymmetric ejecta deposition around the crater rim. Although the analysis of ejecta splashing is an interesting problem, we have not yet analyzed the details of ejecta splashing. That is a possible future work. 

\begin{figure}
\includegraphics[clip,width=1.0\linewidth]{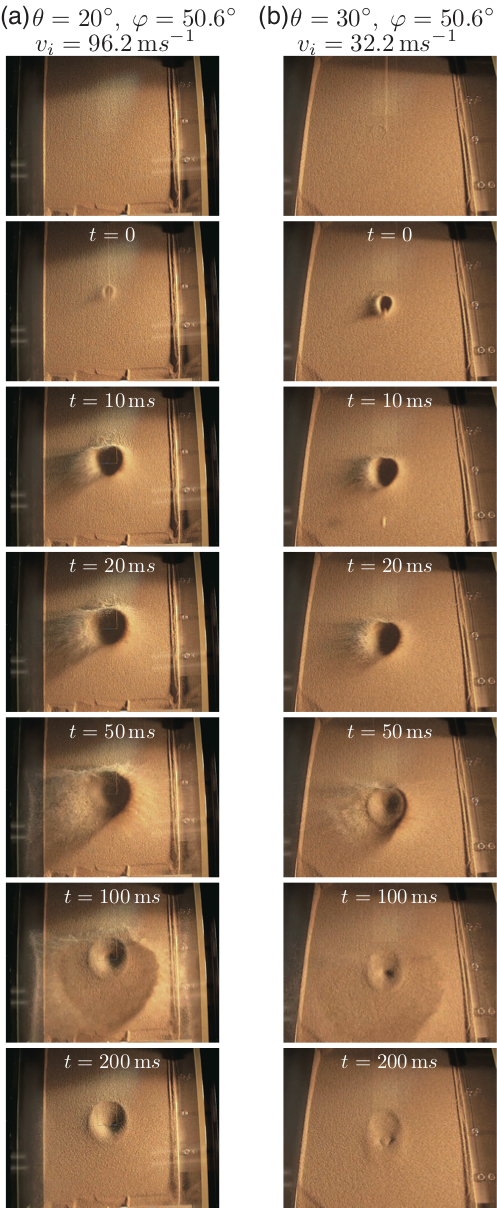}% Here is how to import EPS art
\caption{\label{fig:F5-casio} Topview images of the impacts. Experimental conditions of (a) and (b) are identical to those in Fig.~\ref{fig:F4-photron}(a) and (b), respectively. Although it is difficult to track the projectile motions due to the insufficient frame rate, the asymmetric splashing and crater-wall collapse can be observed in these pictures. These data were used to qualitatively observe the splashing and modification processes.}
\end{figure}

\subsubsection{Topview}
Next, the topview images are shown in Fig.~\ref{fig:F5-casio}, which are the same impact events shown in Fig.~\ref{fig:F4-photron} but observed from different angles. Due to the limitations of shutter speed and temporal resolution, it is difficult to identify the projectiles in Fig.~\ref{fig:F5-casio}. However, the ejecta splashing can clearly be observed. In particular, the asymmetric ejecta deposition can be confirmed around $t=50$--$100$~ms. In addition, the asymmetric crater-wall collapse can be seen at $t=200$~ms, particularly in Fig.~\ref{fig:F5-casio}(b). This crater-wall collapse is caused by the instability of the upper crater wall, where the angle exceeds the repose angle. Since the initial inclination angle $\theta=30^{\circ}$ is very close to the angle of repose $\theta_r=34^{\circ}$, the transient crater-wall angle can easily become greater than $\theta_r$ in the excavation (cavity formation) stage (Fig.~\ref{fig:F5-casio}(b)). Thus, the crater-wall collapse is inevitable in the impact occurring on the steeply inclined granular target. This collapse behavior is consistent with previous experiments~\cite{Hayashi:2017,Aschauer:2017}. However, quantitative image analysis of the topview images shown in Fig.~\ref{fig:F5-casio} is difficult. Therefore, we can only use these images to qualitatively analyze the cratering and the subsequent modification process.

Some of the splashed ejecta seem to reach the container sidewall. Such an effect is almost negligible in this system, because the amount of ejecta splashing far away seems to be minimal. 

\subsection{Crater morphology}
\begin{figure}
\includegraphics[clip,width=1.0\linewidth]{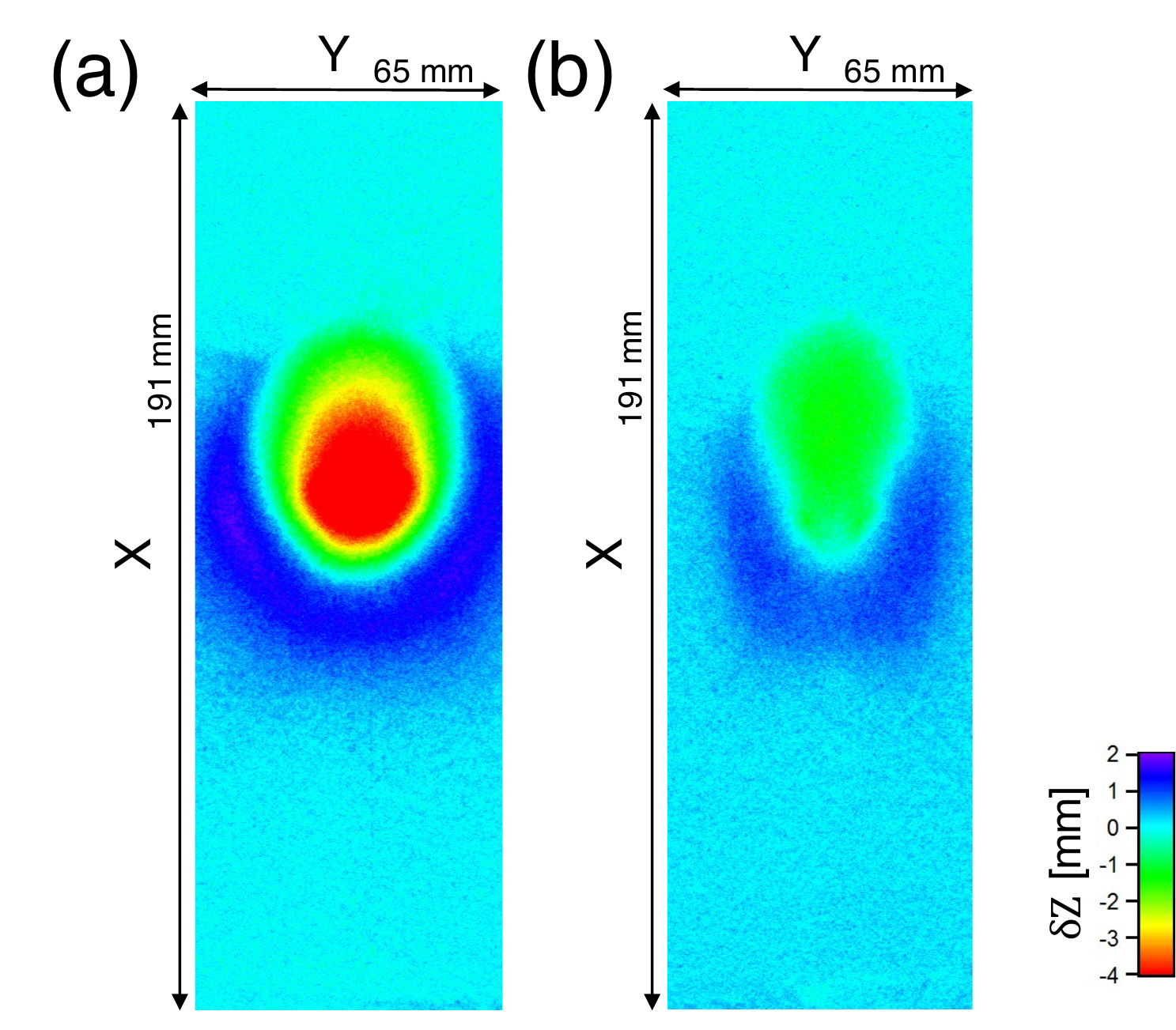}% Here is how to import EPS art
\caption{\label{fig:F6-profile} Crater morphologies. The final crater profiles produced by the impacts of Figs.~\ref{fig:F4-photron} (and \ref{fig:F5-casio}) (a) and (b) respectively, are shown in panels (a) and (b). The surface height $\delta Z$ was computed from the height difference between before and after the impact. The asymmetric ejecta deposition and crater-cavity collapse can be observed. The systematic experiment and detailed analysis of the crater morphologies can be found in Ref.~\onlinecite{Takizawa:2019b}.}
\end{figure}

For the quantitative analysis of crater morphology, the laser profilometry data should be used. In Fig.~\ref{fig:F6-profile}, the measured crater shapes are shown. Experimental conditions for the data shown in Fig.~\ref{fig:F6-profile} are identical to those in Figs.~\ref{fig:F4-photron} and \ref{fig:F5-casio}. The measured crater profiles clearly indicate the asymmetric final cavity shape as well as asymmetric ejecta deposition. The transient crater cavity shape was significantly modified by the crater-wall collapse (landslide), as observed in Fig.~\ref{fig:F6-profile}(b). Using these crater profiles, characteristic length scales of the resultant crater shape can be defined. In Ref.~\onlinecite{Takizawa:2019b}, the width, length, depth, and volume of the crater cavity were analyzed using the systematic dimensional analysis called $\Pi$-groups method~\cite{Holsapple:1993}. From careful dimensional analysis of the resultant crater morphologies, a set of scaling laws for asymmetric craters was obtained~\cite{Takizawa:2019b}. Using the scaling laws, we can discuss a possible way to estimate the origin of peculiar (asymmetric) craters found on various astronomical bodies~\cite{Takizawa:2019b}. As mentioned in Sec.~\ref{sec:rebound}, most of the impacts resulted in a projectile rebound. However, the projectile usually bounced off toward a point far from the initial impact point. In addition, the momentum of the secondary impact was quite small. Thus, we could ignore the effect of the secondary impact in most cases. 

\subsection{Rebound analysis}
\label{sec:rebound}
\begin{figure}
\includegraphics[clip,width=1.0\linewidth]{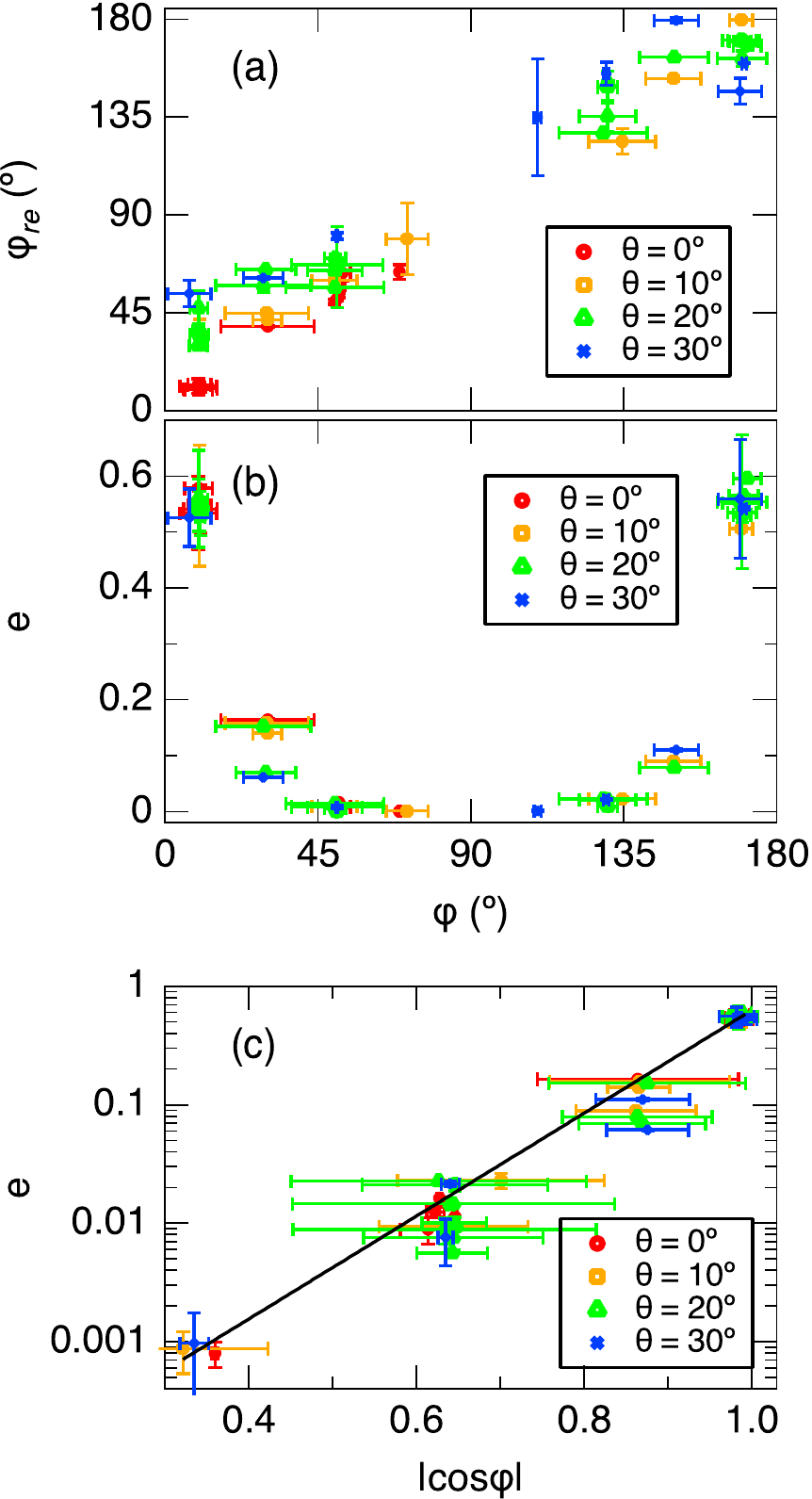}% Here is how to import EPS art
\caption{\label{fig:F7-rebound} Rebound angle and restitution coefficient of the oblique impact onto an inclined granular target. (a) The relationship between impact angle $\varphi$ and rebound angle $\varphi_{\rm re}$. The colors and symbols indicate the inclination angle $\theta$. The simple tendency $\varphi_{\rm re} \simeq \varphi$ can be observed independent of $\theta$. (b) The restitution coefficient $e$ is plotted as a function of incident angle $\varphi$. One can confirm that $e$ is almost independent of $\theta$ and symmetric around the normal impact ($\varphi=90^{\circ}$). (c) The relationship between $e$ and $|\cos\varphi|$ is plotted in the semi-log format. The clear exponential relationship (linear relationship in semi-log plot) can be confirmed. Errors were computed from the image analysis uncertainties and error propagation.}
\end{figure}

Finally, we briefly discuss the rebound dynamics. The relationship between impact angle $\varphi$ and rebound angle $\varphi_{\rm re}$ is plotted in Fig.~\ref{fig:F7-rebound}(a). One can confirm the tendency of coincidence between $\varphi$ and $\varphi_{\rm re}$. The error bars shown in Fig.~\ref{fig:F7-rebound} are computed from the standard deviation of the image analysis (Sec.~\ref{sec:sideview_image}). As seen in Fig.~\ref{fig:F7-rebound}(a), the measurement errors are probably the main source of data scattering. However, we can roughly assume $\varphi \simeq \varphi_{\rm re}$. This means that the energy dissipation is almost independent of direction. In other words, vertical and horizontal components of the restitution coefficient were roughly identical. Thus, we simply defined and measured the restitution coefficient by $e=|v_{\rm re}|/|v_i|$ . Note that the motion in the $Y$ direction was omitted from the calculation of $e$. In Fig.~\ref{fig:F7-rebound}(b), the measured restitution coefficient $e$ as a function of incident angle $\varphi$ for various inclination angle $\theta$ is plotted. As can be seen in Fig.~\ref{fig:F7-rebound}(b), rebounds occurred frequently except at normal incident angles ($\varphi=90^{\circ}$). One can easily confirm that $e$ is almost independent of $\theta$. In addition, the behavior of $e$ in Fig.~\ref{fig:F7-rebound}(b) appears symmetric around $\varphi=90^{\circ}$. This means that the downward impact and upward impact resulted in the same $e$ if the angle from the surface was identical; $e(\varphi) \simeq e(180^{\circ}-\varphi)$. To clearly show this trend, $e$ was plotted as a function of $|\cos\varphi |$ in Fig.~\ref{fig:F7-rebound}(c). As observed, all the data seem to obey a simple exponential form; $e \sim \exp(|\cos\varphi|)$. While this result is not very relevant to planetary impact cratering, such an experimental result could be interesting as a fundamental aspect of the physics of granular impact. That is, the developed system is also useful for studying the physics of granular impact itself.  Furthermore, the granular bouncing dynamics is an important property for spacecraft touchdown missions on the surface of small asteroids. The much more detailed evaluation of the rebound dynamics with various impact angles and terrain slopes is an important future problem.

\section{Discussion}
The developed experimental system can also be used to study another planetary-related problem. That is the relaxation of the large-scale sloped terrain. The relaxation of the large-scale slope can be induced by the accumulation of small-scale impacts on its surface. In general, when the sloped regolith terrain is subjected to vibration or relatively small impacts, a relaxation of the slope can take place. Recently, the vibration-induced slope relaxation has been studied extensively~\cite{Roering:1999,Richardson:2004,Richardson:2005,Tsuji:2018,Tsuji:2019}. In addition, the triggering of avalanching flow by micro impacts has been studied under microgravity conditions~\cite{Hofmann:2017}. When a tiny impactor collides on the surface of a large crater wall, the crater wall is slightly relaxed by the asymmetric ejecta deposition and landsliding. By accumulating this process on a long (astronomical) timescale, the crater shape is degraded. Such crater relaxation process was theoretically studied by Soderblom (1970)~\cite{Soderblom:1970}. This type of relaxation mode becomes dominant on the surface of gravity-dominant bodies such as the Moon and Mars. For small bodies, a global seismic shaking effect~\cite{Richardson:2004,Richardson:2005} becomes dominant. However, to the best of our knowledge, there has not been an experimental study concerning the actual mechanics of the small impact-induced relaxation process for relatively large-gravity bodies. We can investigate the effective relaxation of the sloped target layer by computing the migration of the center of mass caused by the impact cratering on an inclined surface. The detailed analysis of the relaxation dynamics is an ongoing project using this experimental system, and the results will be published elsewhere in near future. Specifically, we can study both macroscopic (crater shape) and microscopic (slope relaxation) dynamics using the experimental setup developed in this study.

\section{Conclusions}
We developed a novel experimental apparatus for the oblique impact of a plastic projectile ($D_i=6$~mm) onto an inclined granular surface. The incident angle $\varphi$ and inclination angle of the target surface $\theta$ can be independently controlled in the system. Natural sand was used for the target to mimic a regolith layer. The barrel of the projectile injection unit was made by 3D printer, and the achievable maximum impact velocity of this injection unit was about $v_i\simeq 100$~m~s$^{-1}$. The impact dynamics was filmed by two high-speed cameras to measure impact parameters or qualitatively observe the cratering process. Before and after the impact, the target surface profile was measured using a laser profiler. From the difference of surface profiles before and after the impact, the final crater shape can be precisely measured. Using the developed system, we can study the macroscopic crater morphology that is affected by the effects of oblique impact and inclination of the target terrain. At the same time, the effective slope relaxation due to the microimpact can also be modeled by this system. Therefore, the developed system enables investigation of both large-scale and small-scale dynamics of cratering and crater degradation. Besides, the fundamental physics of granular impact phenomena can also be examined by using the developed system. For instance, we found that the restitution coefficient was independent of $\theta$ and exponentially increases as $e\sim\exp(|\cos\varphi|)$. We have reported the details on the macroscopic crater morphologies elsewhere~\cite{Takizawa:2019b} and are preparing a paper for the slope relaxation analysis.

%\begin{table}
%\caption{\label{tab:table4}Numbers in columns Three--Five have been
%aligned by using the ``d'' column specifier (requires the
%\texttt{dcolumn} package). 
%Non-numeric entries (those entries without
%a ``.'') in a ``d'' column are aligned on the decimal point. 
%Use the
%``D'' specifier for more complex layouts. }
%\begin{ruledtabular}
%\begin{tabular}{ccddd}
%One&Two&\mbox{Three}&\mbox{Four}&\mbox{Five}\\
%\hline
%one&two&\mbox{three}&\mbox{four}&\mbox{five}\\
%He&2& 2.77234 & 45672. & 0.69 \\
%C\footnote{Some tables require footnotes.}
%  &C\footnote{Some tables need more than one footnote.}
%  & 12537.64 & 37.66345 & 86.37 \\
%\end{tabular}
%\end{ruledtabular}
%\end{table}

\begin{acknowledgments}
This work was supported by JSPS KAKENHI grant No.~18H03679. 
\end{acknowledgments}

\bibliography{ImpactInstrum}% Produces the bibliography via BibTeX.

\end{document}